\documentclass[11pt]{article}

\def\ve{\varepsilon}

\renewcommand{\thefootnote}{\fnsymbol{footnote}}

\begin{document}
\begin{titlepage}


\vspace*{1cm}
\begin{center}
{\LARGE Hierarchy and Wave Functions in a \\ Simple Quantum Cosmology}
\end{center}

\vspace{1cm}
\begin{center}
{\sc T. R. Mongan \footnote{E-mail: tmongan@mail.com}} \\
{\it 84 Marin Avenue\\
Sausalito, CA 94965, U.S.A.}
\end{center}

\vspace{1cm}

\begin{abstract}
\noindent

Astrophysical observations indicate the universe is asymptotic to a de
Sitter space with  vacuum energy density $\Omega_A=0.7$.  Then, since
$H_0\approx65$ km sec$^{-1}$
Mpc$^{-1}$, our universe has  only about $10^{122}$ degrees of freedom.  A
finite number of degrees of freedom is consistent  with a closed
universe arising from a quantum fluctuation, with zero total quantum  
numbers.  If space-time has eleven dimensions, and the universe began
as a closed force-symmetric ten-dimensional space with characteristic
length $l$, seven of the space dimensions must have collapsed to
generate the three large space dimensions we  perceive.  The
holographic entropy conjecture then suggests the initial length scale
$l$ must  be roughly twenty orders of magnitude larger than the Planck
length.  
Accordingly, the nuclear force must be about forty orders of magnitude
stronger than gravity, possibly resolving the force heirarchy problem.
A  wavefunction for the radius of
curvature of the universe can be obtained from the  Schr\"odinger
equation derived by Elbaz and Novello.  The product of this
wavefunction  and its complex conjugate can be interpreted as the
probability density for finding a given  radius of curvature in one of
the infinity of measurements of the radius of curvature  possible (in
principle) at any location in a homogeneous isotropic universe. 
\end{abstract}
\end{titlepage}

\renewcommand{\thefootnote}{\arabic{footnote}}
\setcounter{footnote}{0}
%


The nuclear force is $1.7 \times 10^{38}$ times stronger than gravity,
and there are indications this force hierarchy relates to the vacuum
energy density/cosmological constant.  For instance, if the proton
mass is specified, the observed vacuum energy density can be
calculated from a simple quantum cosmological model \cite{Mongan} of
eleven-dimensional space-time.  In another approach, Mena Marugan and
Carneiro \cite{MC} consider a three-dimensional universe dominated by
the observed vacuum energy density and apply the holographic
conjecture \cite{Bousso} to determine the total number of observable
degrees of freedom in the universe.  They set this equal to the
estimated number of degrees of freedom for elementary excitations of
typical size $l_N$ inside the maximum observable Hubble radius.  They
then estimate the proton mass from the length scale $l_N$, using the
uncertainty principle.  The result is about twenty orders of magnitude
smaller than the Planck mass, suggesting a nuclear force roughly $40$
orders of magnitude stronger than 
gravity.  In contrast, the approach outlined below uses the
holographic conjecture to relate the force hierarchy to the existence
of seven extra space dimensions.  

The Friedmann equation for the radius of curvature $R$ of a closed
homogeneous isotropic universe \cite{Islam,MTW} is 
\[
\dot{R}^2 - \left( {8\pi G\over 3}\right) \left[ \, \varepsilon_r 
\left({R_0\over R}\right)^4 +\, \varepsilon_m \left({R_0\over R}\right)^3 +
\varepsilon_v \right] \left({R\over c}\right)^2 \ = \ -c^2
\]
where $\varepsilon_r,\varepsilon_m,\varepsilon_v$  and $R_0$ are,
respectively, the present values of the radiation, matter and vacuum
energy densities and the radius of curvature.  Astrophysical
measurements indicate the expansion of the universe is accelerating,
and the energy density of the universe is dominated by a cosmological
constant/vacuum energy density with $\Omega_\Lambda=0.7$\,.  The
cosmological constant is related to the vacuum energy density
\cite{Islam} by $\Lambda=8\pi G\ve_v/c^4$.  As $R\rightarrow \infty$,
the radiation and matter energy density (and the curvature
energy) are driven to zero by the expansion of the universe, and the
Friedmann equation becomes  
\[
\left({\dot{R}\over R}\right)^2={8\pi G \ve_v\over c^2} = {\Lambda
  c^2\over 3} \ .
\]
  So, our universe is
asymptotic to a de Sitter space (the vacuum solution to Einstein's
equations with a positive cosmological constant).  There is a
cosmological horizon in a de Sitter space, because no signal can be
received from beyond the Hubble radius where the expansion velocity
equals the speed of light.  The Hubble radius at $\dot{R}=c$, the maximum
observable Hubble distance, is $\sqrt{3/\Lambda}$ and the area of the
cosmological 
horizon is $A=12\pi/\Lambda$.  Then, according to the holographic
conjecture, the 
number of observable degrees of freedom in the universe is
$N=A/4=3\pi/\Lambda$.  For a 
Hubble constant $H_0 = 65$ \mbox{km sec$^{-1}$ Mpc$^{-1}$} \cite{Primack},
the critical density $\rho_c=3H_0^2/8\pi G = 7.9 \times 10^{-30}$ 
\mbox{g cm$^{-3}$}, the vacuum energy density  $\ve_v=0.7\, \rho_c\, c^2=
5.0 \times 10^{-9}$  \mbox{g cm$^2$ sec$^{-2}$ cm$^{-3}$}, and
$\Lambda= 1.0 \times 10^{-56}$  cm$^{-2}$.  So, $N=3\pi/(\Lambda
\delta^2)= 3.5 \times 10^{122}$, where $\delta$ is the Planck length.

A finite number of degrees of freedom is consistent with a closed
universe arising from a quantum fluctuation, with zero total quantum
numbers.  If space-time has eleven dimensions, the simple quantum
cosmology in \cite{Mongan} envisions our universe as a closed
ten-dimensional space that is the direct product of a closed
three-dimensional subspace 
and a closed seven-dimensional subspace as the processes that break
force symmetry  begin.  
Initially, both spaces have length scale $l$ and radius $l/2\pi$, and all
forces have equal strength.  The length scale is related to the
initial strength $G_i$ of the nuclear and gravitational force by
$l=\sqrt{\hbar G_i/c^3}$.

In the model in \cite{Mongan}, the seven-dimensional subspace
collapsed and injected entropy into the three-dimensional subspace,
generating the three large space dimensions we inhabit.  This suggests
that the initial size of the direct product space must be large enough
to contain the number of degrees of freedom needed to describe the
evolution of 
our observable universe.  According to the holographic conjecture, the
number of degrees  of freedom available for describing the light
sheets of any surface is bounded by one quarter of the area $A$ of the
surface in Planck units \cite{Bousso3}.  The characteristic dimension of  
the initial state of the universe can be estimated by extending the
holographic conjecture  to physical systems with more than three space
dimensions. 

In the initial state, the seven-dimensional subspace dominates the
contribution of the three-dimensional subspace to the available
degrees of freedom because of its higher dimensionality.  The area of
a seven-sphere \cite{Bousso2} of radius $l/2\pi$ is ${16\pi^3\over
  15}\left({l\over 2\pi \delta }\right)^6$  in units of the Planck
length $\delta$.
  Denote the initial surface area of the seven-dimensional
subspace by $A_7={16f\pi^3\over 15}\left({l\over 2\pi
    \delta}\right)^6$, where  $f$  is greater than one, to allow for the
fact that the topology of the seven-dimensional subspace is likely to be
considerably more intricate than that of a seven-sphere.  If the
holographic bound is saturated, so $N=A_7/4$, the number of degrees of freedom
in the seven-dimensional subspace is $N={4f\pi^3\over 15}
\left({l\over 2\pi \delta}\right)^6$.  For  
$N = 3.5 \times 10^{122},\, l = (1.2 \times 10^{21})f^{-1/6}\delta$.
This indicates the nuclear force would have to be about forty-two
orders of magnitude stronger than gravity to insure that an initial
state involving a spherical seven-dimensional subspace ($f = 1$) could
contain all the degrees of freedom necessary to describe our
observable universe.  However, candidate extra dimensional spaces in
theories of the fundamental forces governing the universe are far more
complex than a sphere (see, e.g., the representations of Calabi-Yau
manifolds in \cite{Greene}).  Specification of the detailed nature of
the seven extra space dimensions of eleven-dimensional space-time
requires a detailed theory of the fundamental forces (e.g., a finite
dimensional theory asymptotic to M-theory as the number of degrees of
freedom goes to infinity \cite{Bousso2}).  If, for example, the
perimeter of any planar section of the seven-dimensional subspace with
characteristic radius $l/2\pi$ is $90\,l$, the surface area of the
subspace will be $5 \times 10^{11}$ times the area of a seven-sphere
with radius $l/2\pi$ .  In this case, 
in eleven-dimensional space-time, an initial length scale $l = 2.1 \times
10^{-14}$ cm (and a nuclear force $1.7 \times 10^{38}$ times stronger
than gravity) is necessary to produce an initial seven-dimensional subspace
containing enough degrees of freedom to describe our vacuum energy
dominated universe with $\Omega_\Lambda=0.7$\,.  

On another subject, interpreting the wavefunction of the universe can
be problematic in quantum cosmology.  In the quantum cosmology
outlined in \cite{Mongan}, the stationary state wavefunction for the
radius of curvature of our three-dimensional universe is obtained from
the Schr\"odinger equation 
\[
-{\hbar^2\over 2m} {d^2\ \over dR^2}\; \psi - {4\pi m G\over 3}
\left(\ve_m+\ve_r+\ve_v \right) \left( {R\over c}\right)^2 \psi = -E\,
\psi
\]
derived by Elbaz {\it et al} \cite{Elbaz} and Novello {\it et al}
\cite{Novello}. 
The time-dependent wavefunction of the universe is a superposition of
stationary state solutions with energies close to the Einstein energy
$E = -{1\over2} m c^2$.  In principle, the radius of curvature of the universe
can be measured in an infinity of directions at any point in a
homogeneous and isotropic universe.  The product of the time-dependent
wavefunction and its complex conjugate can be interpreted as the
probability density for finding a given value of the radius of
curvature in one of these measurements.

\end{document}